\documentclass[intlimits,twoside,a4paper]{article}

\usepackage{amsmath,amssymb}
\usepackage{graphicx}

\usepackage[T2A]{fontenc}
\usepackage[cp1251]{inputenc}
\usepackage[eqsecnum]{cmpj2}

\newcommand{\rmd}{\mathrm{d}}

\issue{2013}{16}{2}{23605}

\doinumber{10.5488/CMP.16.23605}

\title[Crossover scaling in the two-dimensional three-state Potts model]%
{Crossover scaling in the two-dimensional three-state Potts model}
\author[T. Nagai, Y. Okamoto, W. Janke]{T.~Nagai\refaddr{label1}, Y.~Okamoto\refaddr{label1,label2,label3,label4}, W.~Janke\refaddr{label5,label6}
}

\addresses{
\addr{label1} Department of Physics, Graduate School of Science,
Nagoya University, Nagoya, Aichi 464--8602, Japan
\addr{label2} Structural Biology Research Center, Graduate School of Science, Nagoya University, Nagoya, \\ Aichi 464--8602, Japan
\addr{label3} Center for Computational Science, Graduate School of Engineering, Nagoya University, Nagoya, Aichi \\ 464--8603, Japan
\addr{label4} Information Technology Center, Nagoya University, Nagoya,\\
Aichi 464--8601, Japan
\addr{label5} Institut f\"ur Theoretische Physik, Universit\"at Leipzig,
Postfach 100 920, 04009 Leipzig, Germany
\addr{label6} Centre for Theoretical Sciences (NTZ), Universit\"at Leipzig,
Postfach 100 920, 04009 Leipzig, Germany
}

\date{Received February 26, 2013, in final form April 19, 2013}

\begin{document}

\maketitle

\begin{abstract}
We apply simulated tempering and magnetizing (STM) Monte Carlo simulations
to the two-dimensional three-state Potts model in an external magnetic field
in order to investigate the crossover scaling behaviour in the temperature-field
plane at the Potts critical point and towards the Ising universality class for
negative magnetic fields. Our data set has been generated by STM simulations
of several square lattices with sizes up to $160\times 160$ spins, supplemented by
conventional canonical simulations of larger lattices at selected simulation
points. We present careful scaling and finite-size scaling analyses of
the crossover behaviour with respect to temperature, magnetic field and
lattice size.
\pacs 64.60.De,75.30.Kz,75.10.Hk,05.10.Ln
\end{abstract}

\section{Introduction}

The two-dimensional three-state Potts model in an external magnetic
field \cite{Potts1952,Wu1982} has several interesting applications in
condensed matter physics \cite{Wu1982}, and its three-dimensional
counterpart serves as an effective model for quantum
chromodynamics \cite{philipsen2006qcd,kim20053,Karsch2000,Mercado2012}.
When one of the three states per spin is disfavoured in an external (negative)
magnetic field, the other two states exhibit $Z_2$ symmetry and one expects
a crossover from Potts to Ising critical behaviour. In the vicinity of the
Potts critical point, another crossover effect takes place when approaching
the critical point along different paths in the temperature-field plane.

To cover such a two-dimensional parameter space, generalized-ensemble Monte
Carlo simulations are useful tools \cite{Janke1998,Hansmann1999,Mitsutake2001,janke2008rugged}. Well-known examples are the
multicanonical (MUCA) algorithm
\cite{berg1991multicanonical,berg1992multicanonical},
the closely related Wang-Landau method \cite{Wang2001a,Wang2001b},
the replica-exchange method (REM) \cite{hukushima1996exchange,Geyer1991}
(see also \cite{Swendsen1986,wang2004replica}), also
often referred to as parallel tempering, and
simulated tempering (ST) \cite{Lyubartsev1992,marinari1992simulated}.
Inspired by recent multi-dimensional generalizations of generalized-ensemble
algorithms \cite{Mitsutake2009multidimensional1,Mitsutake2009multidimensional2,Mitsutake2009MSTMREM},
the ``Simulated Tempering and Magnetizing'' (STM) method has been proposed
by two of us and tested for the classical Ising model
in an external magnetic field \cite{Nagai2012Proc,Nagai2012inpress}. Recently,
we have extended this new simulation method to the two-dimensional
three-state Potts model and by this means generated accurate
numerical data in the temperature-field plane \cite{Nagai2013}.
Here we focus on a discussion of the two above mentioned crossover-scaling scenarios that includes an analysis of the specific heat which especially
   for the Potts-to-Ising crossover provides the clearest signals.

The rest of this article is organized as follows. In section 2 we briefly
discuss the model and review the STM method.
In section 3 we present the results of our crossover-scaling analyses
at the phase transitions with respect to temperature, magnetic field and
lattice size. Finally, section 4 contains our conclusions and an outlook
to the future work.

\section{Model and simulation method}

The two-dimensional three-state Potts model in an external magnetic field
is defined through the Hamiltonian
\begin{align}
H&=E-hM\,, \label{eq:H}\\[2.5ex]
E&=-\sum_{\left< i,j\right>} \delta_{\sigma_i, \sigma_j} \label{eq:E}\,,\\
M&=\sum_{i=1}^N \delta_{0, \sigma_i} \,,\label{eq:M}
\end{align}
where $N=L^2$ denotes the total number of spins $\sigma_i \in \{0,1,2\}$ arranged
on the sites of a square $L\times L$ lattice with periodic boundary
conditions, $\delta$ is the Kronecker delta function and $h$ is the external
magnetic field. The sum in (\ref{eq:E}) runs over all nearest-neighbour pairs.
Note that the magnetization $M$ defined in (\ref{eq:M}) takes on the value $M = N$
for the ordered state in 0-direction, $M = 0$ for the ordered states in
1- or 2-direction, and $M = N/3$ in the disordered phase.

\begin{figure}[h]
  \begin{center}
     \includegraphics[width=8cm]{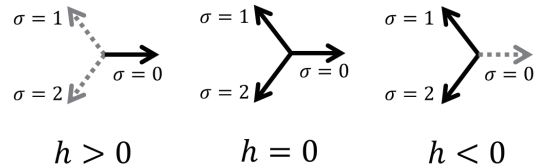}
\caption{\label{fig:spin}
Schematic sketch illustrating the behaviour of the spins of the three-state Potts
model in an external magnetic field $h$. For $h > 0$, the spin state 0 is favoured,
whereas the states 1 and 2 are disfavoured. For $h = 0$, all three states
are equivalent. For $h < 0$, the spin state 0 is disfavoured and the states 1 and 2
are related to each other by $Z_2$ symmetry.
}
  \end{center}
\end{figure}

By mapping the integer valued spins $\sigma_i$ to spin vectors
$\vec{s}_i = \left(\cos(2 \pi \sigma_i/3),\sin(2 \pi \sigma_i/3) \right)$
one readily sees that
$E = (2/3) (-\sum_{\langle i,j \rangle} \vec{s}_i \vec{s}_j - N)$ and
$M = (2/3) (M^{(x)} + N/2)$,
where $M^{(x)}$ is the component of the magnetization vector
$\vec{M} = \sum_i \vec{s}_i$
in field direction (assumed to be along the $x$-axis). In this
equivalent notation, it is fairly obvious that the $Z_3$ symmetry for $h = 0$ is
broken to $Z_2$ for negative external magnetic fields (see figure~\ref{fig:spin}).

Another frequently employed definition of the magnetization is the so-called
``maximum definition''
\begin{eqnarray}
M_{\mathrm{max}} = N m_{\mathrm{max}}
\equiv \frac{3}{2} \left\{ \max_{j=0,1,2} \left[\sum_{i=1}^{N} \delta_{j,\sigma_i}\right] -\frac{N}{3}\right\},
\end{eqnarray}
which yields the physically more intuitive value of 1
when the system is in one of the three ordered phases and
0 in the disordered phase, respectively.

Let us now turn to a brief description of the employed Monte Carlo simulation method.
In the conventional ST scheme \cite{Lyubartsev1992,marinari1992simulated},
the temperature is considered as an additional dynamical variable besides the
spin degrees of freedom. The STM method is a generalization to a two-dimensional
parameter space where the magnetic field is treated as the second additional
dynamical variable similar to the
temperature \cite{Nagai2012Proc,Nagai2012inpress,Nagai2013}. Here, one considers
\begin{align}
\re^{-(E-hM)/T +a(T,h)} \label{eq:f1}
\end{align}
as a joint probability for $(x,T,h)$ ($\in X \otimes \{ T_1, T_2, \cdots, T_{N_T}\} \otimes \{h_1, h_2, \cdots, h_{N_h}\}$), where
$a(T,h)$ is a parameter, $x$ denotes a (microscopic) state,
and $X$ is the sampling space. We have set Boltzmann's constant to unity.
Note that the temperature and external field are discretized into $N_T$ and $N_h$
values, respectively.

\begin{figure}[ht]
   \begin{center}
       \includegraphics[width=5.0cm, clip , angle = 270]{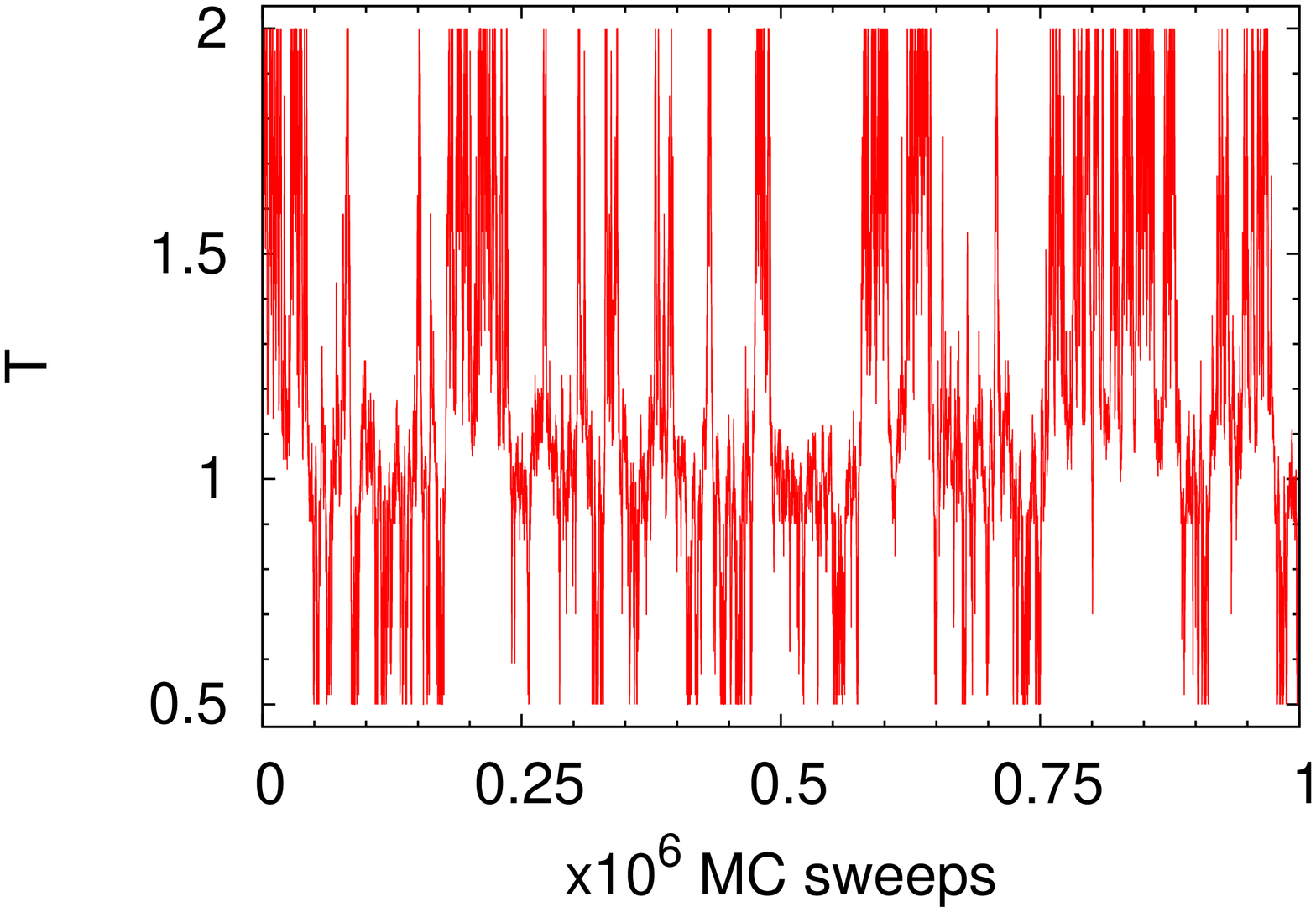}
       \hfill
       \includegraphics[width=5.0cm, clip , angle = 270]{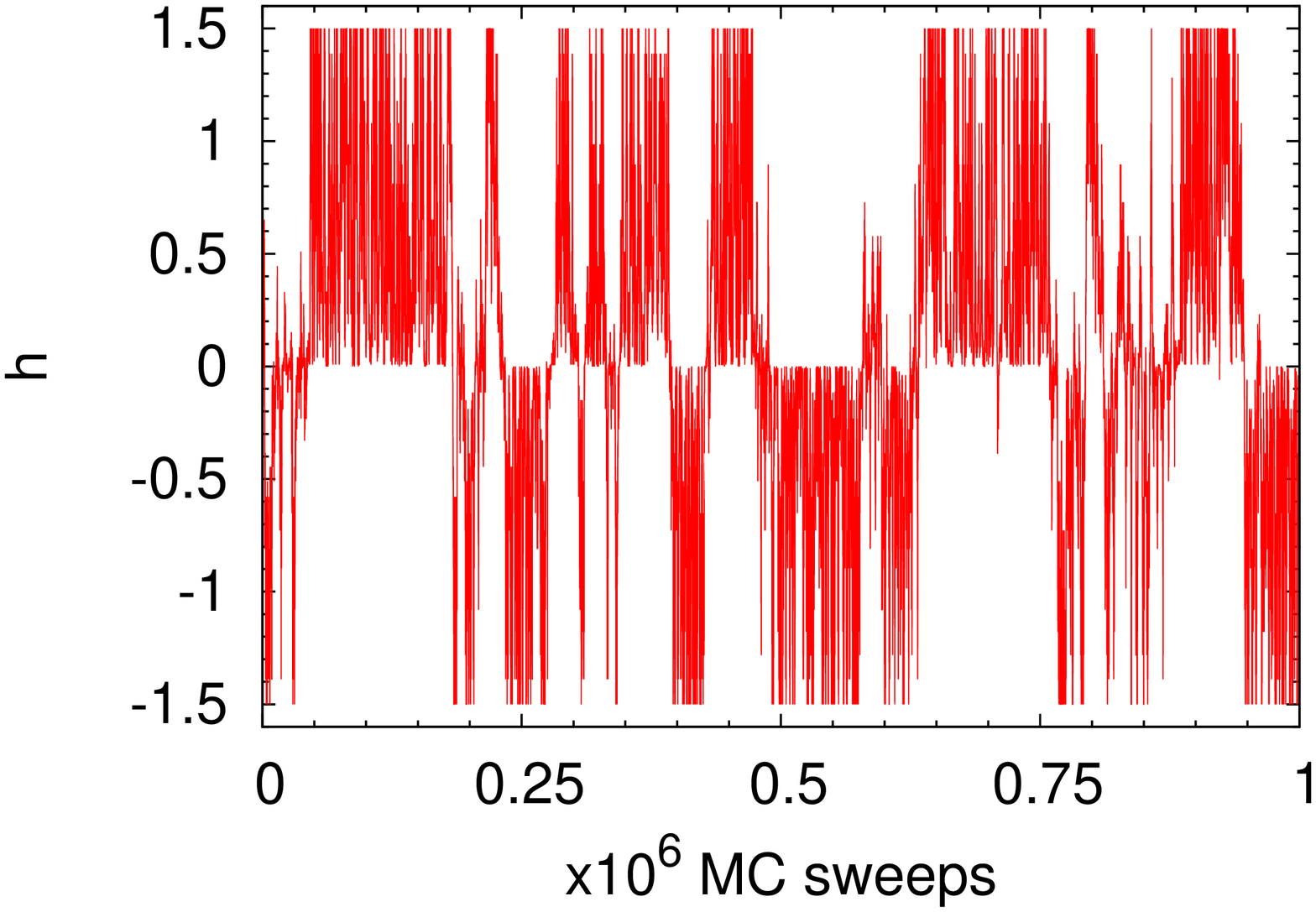}
       \hspace{5mm}
       \caption{(Color online) Time history of temperature $T$ and magnetic field $h$ in
       STM simulations for the linear lattice size~$L=80$.}
       \label{fig:t_extf}
   \end{center}
\end{figure}

A suitable candidate for $a(T_i,h_j)$ can be obtained from
the (empirical) probability of occupying each set of parameter values,
\begin{align}
P(T_i,h_j) = \re^{-f(T_k,h_l) +a(T_k, h_l)} \,,
\end{align}
where $\re^{-f(T_k,h_l)} = \int \rmd x \,\re^{-(E-h_jM)/T_i}$. This shows that
the dimensionless free energy $f(T_i,h_j)$ is the proper choice
for $a(T_i,h_j)$ in order to generate a uniform distribution of the number
of samples according to $T$ and $h$. This implies a random-walk-like
evolution of $T$ and $h$ in STM simulations as it is demonstrated
in figure~\ref{fig:t_extf} for a $80 \times 80$ lattice. The block structures
reflect the first-order phase transition line at $h=0$ in the Potts model
and the second-order phase transition at the effective Ising transition
temperature $T_\mathrm{c} \approx 1.1346$ for negative magnetic field.

\section{Results}

Our STM simulations were performed for lattice sizes $L = 5, 10, 20, 40, 80$,
and 160 with the total number of sweeps varying between about $160 \times 10^6$
and $500 \times 10^6$, with a sweep consisting of $N$ single-spin updates with
the heat-bath algorithm followed by an update of either the temperature $T$ or
the field $h$. We used the Mersenne
Twister \cite{matsumoto1998mersenne} as quasi-random-number generator. Statistical error bars were estimated
using the jackknife blocking
method \cite{Miller1974, efron1982jackknife, berg2004book, Janke2008}.
\begin{figure}[!h]
   \begin{center}
       \includegraphics[width=10cm]{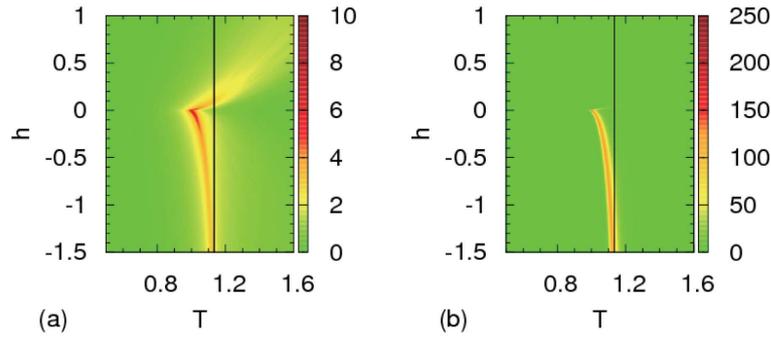}
       \caption{(Color online) (a) Specific heat per site $C/L^2$ and (b) magnetic
       susceptibility per site $\chi /L^2$ as functions of $T$ and $h$
       for $L=80$. The solid vertical line corresponds to $T=1.1346$, which
       is the critical temperature of the Ising model (in 2-state Potts model
       normalization). }
       \label{fig:Ckaimap}
   \end{center}
\end{figure}

Due to its random-walk-like nature, the STM method, combined with
reweighting techniques such as
WHAM \cite{Ferrenberg1989,kumar1992weighted,kumar1995multidimensional} or
MBAR \cite{shirts2008statistically},  yields the density of states $n(E,M)$ (up to
an overall constant) in a wide range of the two-dimensional parameter space.
Using these data it is straightforward to compute a two-dimensional map of
any thermodynamic quantity that can be expressed in terms of $E$ and $M$.
As an example, figure~\ref{fig:Ckaimap}
shows the specific heat $C = (\langle E^2 \rangle - \langle E \rangle^2)/T^2$
and susceptibility
$\chi = (\langle M_{\rm max}^2 \rangle - \langle M_{\rm max} \rangle^2)/T$
per spin as functions of $T$ and $h$ for $L=80$. We see a line of phase
transitions starting at the Potts critical point at $h=0$,
$T_c^{\rm Potts} = 1/\ln(1+\sqrt{3}) = 0.9950$ which, for strong negative magnetic fields,
approaches the
Ising model limit with the critical point at $h \rightarrow -\infty$,
$T_c^{\rm Ising} = 1/\ln(1+\sqrt{2}) = 1.1346$.
For all $h < 0$, the $Z_3$ symmetry of the 3-state Potts model in zero field
is broken to $Z_2$ symmetry (recall figure~\ref{fig:spin}) and by universality
the critical behaviour is expected to be Ising-like.

\begin{figure}[ht]
    \begin{center}
        \includegraphics[width=11.0cm]{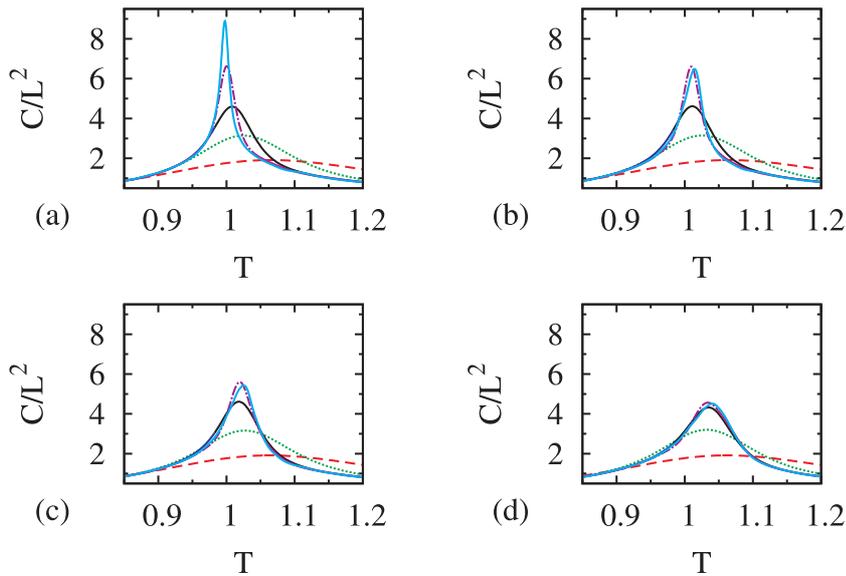}
\caption{(Color online) Specific heat per site $C/L^2$ as a function of $T$. With increasing system size
$L$, the  peaks become more pronounced [$L=5$ (dashed red), $10$ (dotted green), $20$ (solid black), $40$ (dash dotted purple), and $80$ (solid blue)].
                     (a) $h=0.0$, (b) $h=0.005$, (c) $h=0.01$, (d) $h=0.02$. }
        \label{fig:C_merged}
    \end{center}
\end{figure}

For positive magnetic fields, the phase transition disappears altogether.
However, for finite lattices and small $h > 0$, the singular behaviour
persists to some extent due to finite-size effects. More precisely, the peaks
of, e.g., the specific heat shown in figure~\ref{fig:C_merged},  grow with an
increasing lattice size $L$ until $L$ is larger than the (finite) correlation
length of the system. This can be interpreted as a crossover in the dependence of
field $h$ and lattice size $L$.

\begin{table}[h]
\caption{Critical exponents for the
two-dimensional Ising  and three-state Potts models [$y_t=1/\nu$, $y_h=(\beta+\gamma)/\nu$] \cite{Wu1982}.}
\label{tab:expo}
\begin{center}
\begin{tabular}{@{}|l|c|c|c|c|c|c|c|}
\hline
Model  & $y_t$& $y_h$&   $\alpha$ & $\beta$ &  $\gamma $&  $\delta$ & $\nu$  \\
\hline\hline
Ising    &   1&15/8& 0 ({\rm log}) & 1/8    &   7/4      &   15       &1\\
Potts  & 6/5&28/15& 1/3 & 1/9 &  13/9       &  14       &5/6\\
\hline
\end{tabular}
\end{center}
\end{table}

\begin{figure}[ht]
    \begin{center}
        \includegraphics[width=11cm]{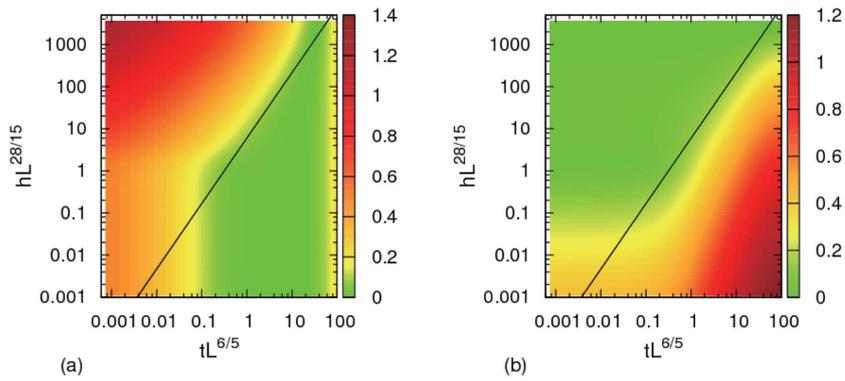}
        \caption{(Color online) The difference between magnetization and its expected scaling behaviours around the critical point for $L=80$.
Shown are
                (a) $|mL^{2/15}-1.2(L^{6/5}t)^{1/9}|$
where the amplitude $1.2$ was obtained by fitting the magnetization data to $t^{1/9}$
and               (b) $|mL^{2/15}-(L^{28/15}h)^{1/14}|$.
                 In both plots, the solid line corresponds to $h=6t^{14/9}$.
                                   }
        \label{fig:3d_scaledMdiff}
    \end{center}
\end{figure}

To study, in the vicinity of the Potts critical point, the crossover-scaling
behaviour in the $T-h$ plane, we calculated the magnetization $m = M/L^2$
by reweighting. Its scaling form is given by \cite{fisher1974renormalization}
\begin{align}
m(T,h,L) = L^{-\beta/\nu} \Psi (tL^{y_t}, hL^{y_h}) \,,
\end{align}
where $y_t = 1/\nu$ and $y_h = (\beta+\gamma)/\nu$ are the usual scaling
dimensions which can be expressed in terms of standard critical exponents.
For easier reference, we have collected the exactly known critical exponents
of the two-dimensional Ising and Potts models in table~\ref{tab:expo}.
The actually observed exponent depends on the precise path in which
the critical point is approached in the $T-h$ plane.
According to the crossover scaling formalism \cite{fisher1974renormalization}
in the limit of an infinite lattice,
if $t^{-y_h/y_t}h$ (in the Potts model $t^{-14/9}h$) is small enough, then the
magnetization obeys $m \sim t^{\beta}$ ($= t^{1/9}$), and
otherwise it scales as $m \sim h^{1/\delta}$ ($= h^{1/14}$),
where $t= (T_\mathrm{c}-T)/T_\mathrm{c}$.
Figure~\ref{fig:3d_scaledMdiff}~(a) shows that as long as finite-size effects
are negligible ($L^{6/5}t\gg0.1$) and $t\gg(h/6)^{9/14}$
(i.e., $t^{-14/9}h$ is small),
then the critical behaviour is $m\sim t^{1/9}$.
Figure~\ref{fig:3d_scaledMdiff}~(b) shows that if finite-size effects are negligible
($L^{28/15}h\gg0.1$) and $t\ll(h/6)^{9/14}$ (i.e., $t^{-14/9}h$ is large),
then the critical behaviour is $m\sim h^{1/14}$.
Thus, figure~\ref{fig:3d_scaledMdiff} clearly demonstrates that the line
$h=6 t^{14/9}$ gives the boundary of the two scaling regimes.

\begin{figure}[ht]
    \begin{center}
    \hspace{3mm}
        \includegraphics[height=5cm,width=7cm]{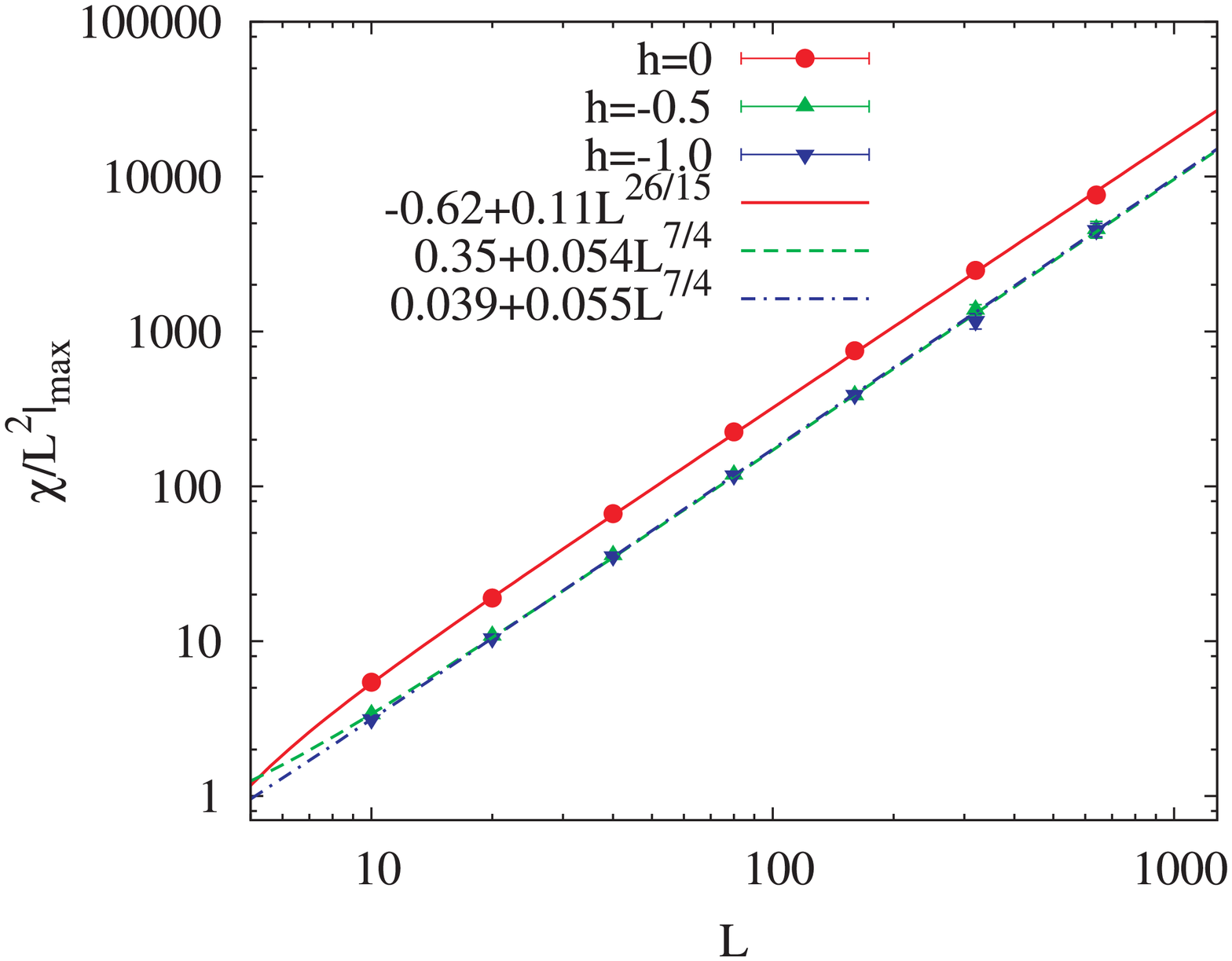}
        \hfill
        \includegraphics[height=5cm,width=7cm]{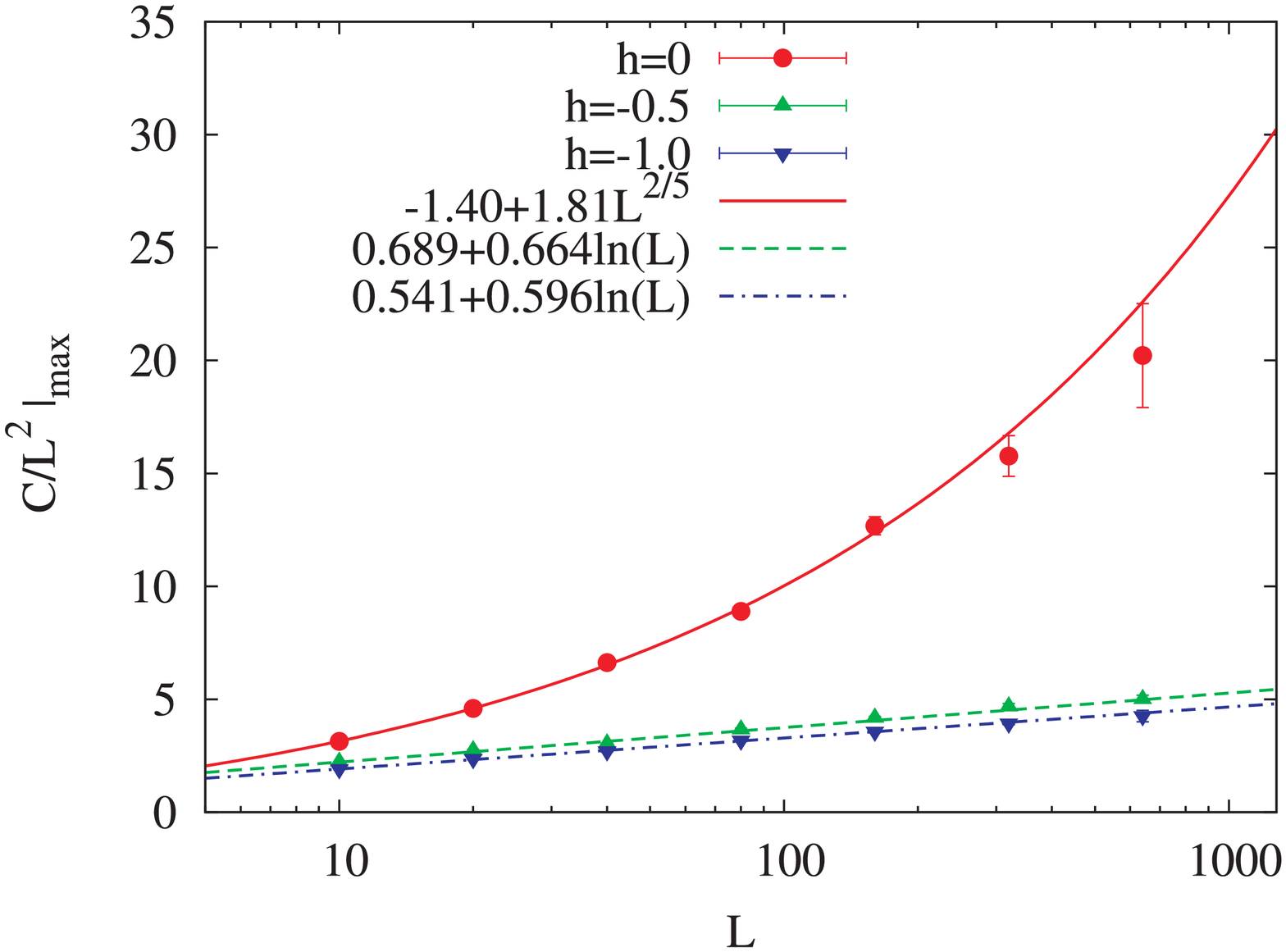}
        \\[2ex]
        \hspace{3mm}
\includegraphics[height=5cm,width=7cm]{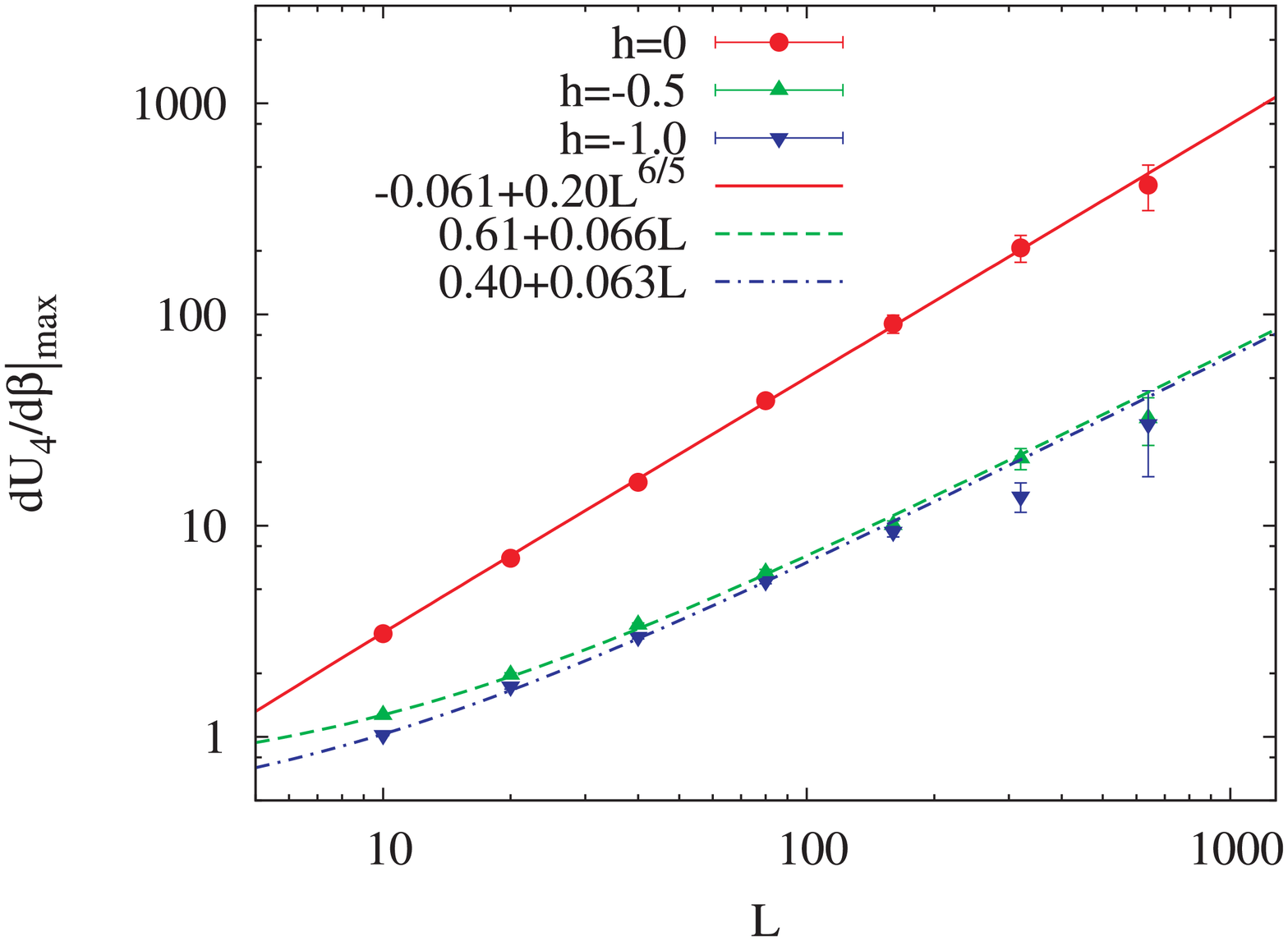}
    \hfill
    \includegraphics[height=5cm,width=7.3cm]{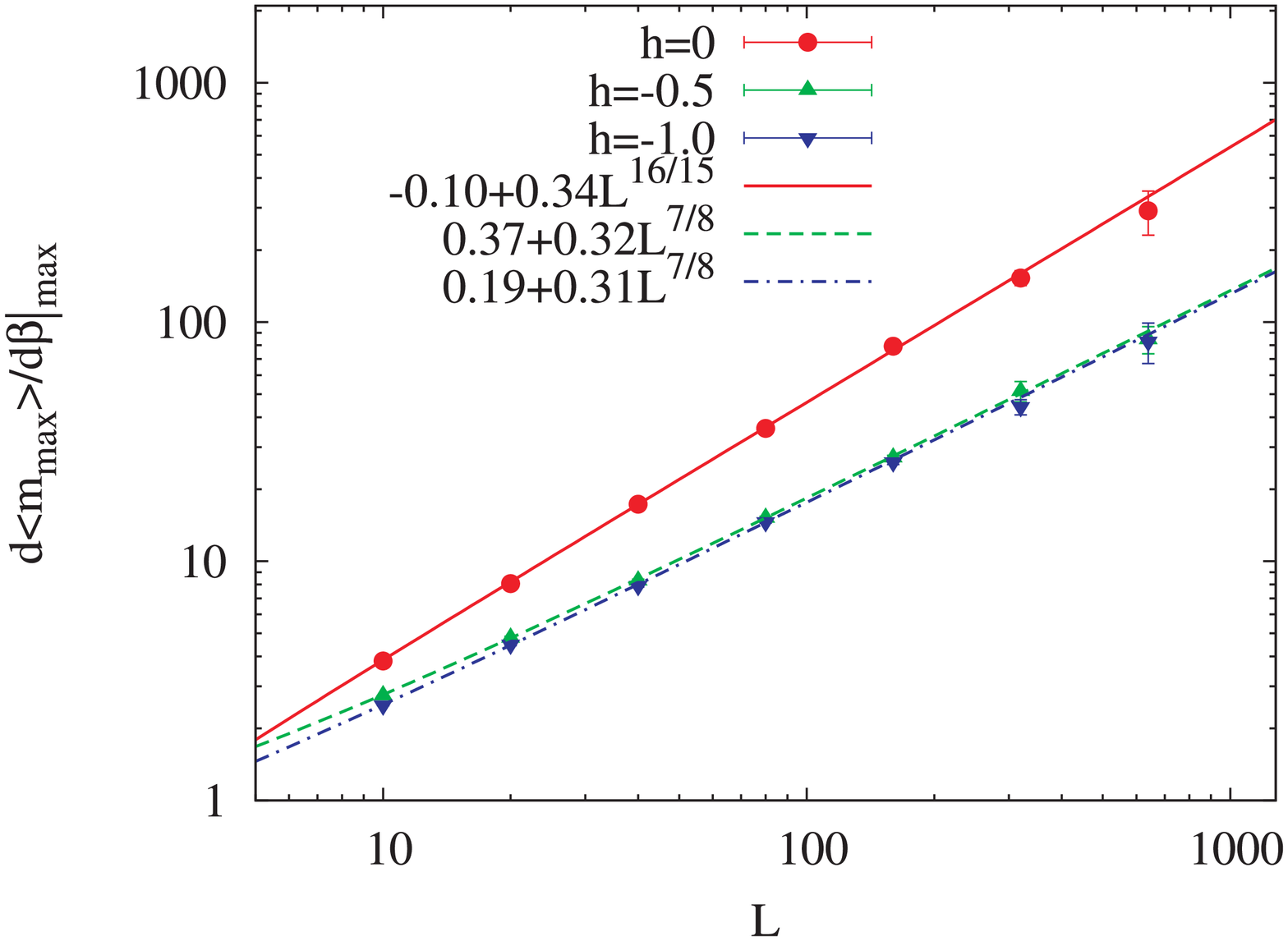}
        \caption{(Color online)
Finite-size scaling behaviour of $\chi_{\mathrm{max}}$, $C_{\mathrm{max}}$,
$\frac{\rmd\ln\left<U_4\right>}{\rmd\beta}|_{\mathrm{max}}$, and
$\frac{\rmd\left<m_\mathrm{max}\right>}{\rmd\beta}|_{\mathrm{max}}$
for three characteristic $h$ values.}
        \label{fig:kaimax}
    \end{center}
\end{figure}

Since the three-state Potts model in a negative magnetic field is expected
to behave like the Ising model, we also investigated the crossover behaviour
between these two models using finite-size scaling techniques. For the
susceptibility maximum
$\chi/L^2|_{\rm max} \propto L^{\gamma/\nu}$,
the finite-size scaling exponents of the Potts and Ising model are given
by $\gamma/\nu={26/15}=1.7333\dots$ and $7/4=1.75$, respectively.
Figure \ref{fig:kaimax} shows that the exponents are so similar that we
can hardly distinguish the difference, despite the accuracy of the measurements.
The difference is much more pronounced for the maxima of the specific heat
which are expected to scale with the system size $L$ with an exponent
$\alpha/\nu = 2/5$ for the Potts and $\alpha/\nu = 0$, i.e., logarithmically,
for the Ising model.
We also measured different quantities, which are the maximum values of
$\frac{\rmd\ln\left<m_\mathrm{max}\right>}{\rmd\beta}$,
$\frac{\rmd\ln\left<m_\mathrm{max}^2\right>}{\rmd\beta}$,
$\frac{\rmd\ln\left<U_2\right>}{\rmd\beta}$,
$\frac{\rmd\ln\left<U_4\right>}{\rmd\beta}$, and
$\frac{\rmd\left<m_\mathrm{max}\right>}{\rmd\beta}$.
Here, $U_2=1-\frac{\left<m^2_{\mathrm{max}} \right>}{3\left<m_{\mathrm{max}} \right>^2}$ and
$U_4=1-\frac{\left<m^4_{\mathrm{max}} \right>}{3\left<m^2_{\mathrm{max}} \right>^2}$ are the Binder cumulants \cite{binder1981finite}.
The derivatives were obtained by using \cite{r355}
\begin{align}\noindent
\frac{\rmd\ln\left<m^k_\mathrm{max}\right>}{\rmd\beta} &=  \left<E\right>-   \frac{\left<m^k_\mathrm{max} E\right>}{\left<m^k_\mathrm{max} \right>} \,, \\
%
%
\frac{\rmd\ln\left<U_{2k}\right>}{\rmd\beta} &=\frac{\left<m_\mathrm{max}^{2k}\right>}{3\left<m^k_\mathrm{max}\right>^2}
\left\{
\left< E\right> -2\frac{\left<m^k_\mathrm{max} E\right>}{\left< m^k_\mathrm{max}\right>}
+ \frac{\left<m^{2k}_\mathrm{max} E\right>}{\left<m^{2k}_\mathrm{max}\right>}
\right\} \,, \\
%
%
\frac{\rmd\left<m_\mathrm{max}\right>}{\rmd\beta} &= \left<m_\mathrm{max} \right> \left<E\right>- \left<m_\mathrm{max} E\right> \,.
\end{align}

Figure \ref{fig:kaimax} shows our results.
Note that
$\frac{\rmd\ln\left<m_\mathrm{max}\right>}{\rmd\beta}|_{\mathrm{max}}$,
$\frac{\rmd\ln\left<m_\mathrm{max}^2\right>}{\rmd\beta}|_{\mathrm{max}}$,
$\frac{\rmd\ln\left<U_2\right>}{\rmd\beta}|_{\mathrm{max}}$,
$\frac{\rmd\ln\left<U_4\right>}{\rmd\beta}|_{\mathrm{max}}$, and \linebreak
$\frac{\rmd\left<m_\mathrm{max}\right>}{\rmd\beta}|_{\mathrm{max}}$
are expected to behave asymptotically as
$L^{1/\nu}$,
$L^{1/\nu}$,
$L^{1/\nu}$,
$L^{1/\nu}$, and
$L^{(1-\beta)/\nu}$,
respectively, as the lattice size $L$ increases \cite{Janke2008}.
These critical exponents are given for the Potts model by
$1/\nu = 6/5$ and $(1-\beta)/\nu = 16/15$,
and for the Ising model by $1/\nu = 1$ and $(1-\beta)/\nu = 7/8$ (see table~\ref{tab:expo}).
We observe that all quantities for $h=0$ (red curve with filled circles)
follow the Potts case and that those for negative external field
(green curve with filled up triangles and blue curve with
filled down triangles)
follow the Ising case in the limit of large $L$. In fact, the two curves
for $h=-0.5$ and $h=-1.0$ converge into almost the same line as
$L$ increases. On the other hand, the (green) curve for $h=-0.5$ exhibits greater deviation from the scaling behaviour for small $L$.
This can also be understood as another crossover effect governed by $h$ and $L$.

\section{Conclusions}

In this work, we reported the scaling and finite-size scaling analyses
of the two-dimensional three-state Potts model in a magnetic field
based on the data generated using the Simulated Tempering and Magnetizing (STM)
method \cite{Nagai2012Proc, Nagai2012inpress}. In such simulations, the
random walk in temperature and magnetic field covers a wide range of
these parameters so that STM simulations enable one to study crossover
phenomena with a single simulation run \cite{Nagai2013}.

By this means we calculated the magnetization, susceptibility, energy,
specific heat and related quantities as functions of temperature, magnetic
field, and lattice size around the critical point using reweighting techniques.
These data allowed us to extract the crossover behaviours of phase
transitions. First, at the Potts critical point for $h = 0$, we observed a
clear crossover of the scaling behaviour of the magnetization with respect
to temperature and magnetic field. Second, from an analysis of the specific
heat and other quantities, a crossover in the scaling laws with respect to
(negative) magnetic field and lattice size was identified, thereby verifying
the expected crossover from 3-state Potts to Ising critical behaviour.

The data of the present work yield the two-dimensional density of states $n(E,M)$
(up to an overall constant) which determines the weight factor for two-dimensional
multicanonical simulations. We hence can also perform two-dimensional multicanonical
simulations, which will be an interesting future task.

As a final remark we should like to stress that the present method is useful not only
for spin systems as considered here but also for other complex systems with
many degrees of freedom. Since our method does not require any change of the
frequent, rather intricate energy calculations, it should be highly compatible
with the available program packages.

\subsection*{Acknowledgements}

We thank
the Information Technology Center, Nagoya University,
the Research Center for Computational Science, Institute for Molecular Science,
and the Supercomputer Center, Institute for Solid State Physics, University of Tokyo,
for computing time on their supercomputers.
This work was supported, in part, by JSPS Institutional Program for Young Researcher Overseas Visit (to T.N.) and
by Grants-in-Aid for Scientific Research
on Innovative Areas (``Fluctuations and Biological Functions")
and for the Computational
Materials Science Initiative from the Ministry of Education,
Culture, Sports, Science and Technology, Japan
(MEXT).
W.J.\ gratefully acknowledges support by DFG Sonderforschungs\-bereich SFB/TRR 102
(Project B04) and the Deutsch-Franz\"osische Hochschule (DFH-UFA) under Grant No.~CDFA--02--07.
T.N.\ also thanks
the Nagoya University Program for ``Leading Graduate Schools: Integrative Graduate Education and Research Program in Green Natural Sciences''
for support of his extended stay in Leipzig.

%
%

\ukrainianpart

\title{Скейлінґ кросоверу у двовимірній тристановій моделі Поттса}
\author{Т. Нагаі\refaddr{label1}, Ю.~Окамото\refaddr{label1,label2,label3,label4}, В.~Янке\refaddr{label5,label6}
}

\addresses{
\addr{label1} Відділ фізики, університет м. Нагоя,  Нагоя, Айчі
464--8602, Японія
\addr{label2} Центр досліджень структурної
біології, університет м. Нагоя,  Нагоя, Айчі 464--8602, Японія
\addr{label3} Центр комп'ютерних наук, університет м. Нагоя, Нагоя,
Айчі 464--8602, Японія
\addr{label4} Центр інформаційних технологій,
університет м. Нагоя,  Нагоя, Айчі 464--8602, Японія
\addr{label5} Інститут теоретичної фізики, університет  Ляйпцігу, 04009 м.
Ляйпціг, Німеччина
\addr{label6} Центр теоретичних природничих наук
(NTZ), університет  Ляйпцігу, 04009  Ляйпціг, Німеччина
}

\makeukrtitle

\begin{abstract}
\tolerance=3000%
Ми застосовуємо Монте Карло симуляції з  симульованим темперуванням
і намагніченням (STM) до двовимірної тристанової моделі Поттса у
зовнішньому магнітному полі для того, щоб дослідити кросоверну
скейлінгову поведінку у площині температура-поле при критичній точці
Поттса, а також  клас універсальності моделі Ізинга для негативних
магнітних полів. Набір наших даних був згенерований STM симуляціями
декількох квадратних ґраток розміром до $160\times 160$ спінів,
доповненими звичайними канонічними симуляціями більших ґраток при
вибраних симуляційних точках. Ми представляємо ретельний аналіз
скейлінгу і скінченомірного скейлінгу кросоверної поведінки по
відношенню до температури, магнітного поля і розміру ґратки.
\keywords тристанова модель Поттса, фазові переходи, критичні
явища, скейлінг кросоверу, симуляції Монте Карло, симульоване
темперування і намагнічення (STM)

\end{abstract}

\end{document}